# Empirical Validation of the Thermal Model of a Passive Solar Cell test

Thierry Alex MARA, François GARDE, Harry BOYER and Malik MAMODE

*University of La Réunion Island, Laboratoire de Génie Industriel, BP 7151, 15 avenue René Cassin, 97 715 Saint-Denis, France. Phone (+262) 93 82 24, Fax (+262) 93 86 65 email : mara@univ-reunion.fr.*

**Abstract**

The paper deals with an empirical validation of a building thermal model. We put the emphasis on sensitivity analysis and on research of inputs/residual correlation to improve our model. In this article, we apply a sensitivity analysis technique in the frequency domain to point out the more important parameters of the model. Then, we compare measured and predicted data of indoor dry-air temperature. When the model is not accurate enough, recourse to time-frequency analysis is of great help to identify the inputs responsible for the major part of error. In our approach, two samples of experimental data are required. The first one is used to calibrate our model the second one to really validate the optimized model.

## 1. Introduction

The improvement of building thermal behaviour is a very important challenge because of the worldwide energy needs. The use of building thermal simulation software is necessary to achieve this task. But, before using such a program, one must ensure that its results are reliable. To do so, a methodology of validation must be applied including the verification of numerical implementation and experimental validation. At University of La Réunion Island, we have developed a dynamic building thermal simulation software called CODYRUN (see ref. [1], for more details). Recently, we checked the software programming using an inter-program comparison procedure called BESTEST (cf. ref. [2]). In this article, we present the empirical validation of the thermal model of a real cell test built with our dynamic simulation program.

Empirical validation is a very important stage in the methodology of validation. However, when comparison between measured and predicted data fails, it's interesting to search for the causes of disagreement which principally derive from :

(1) error measurements;

(2) underestimation of the actual value of one or more parameters;

(3) an error in the model (assumptions, programming error , …).

Parametric sensitivity analysis of the studied model should allow to check point (2). Moreover, this analysis is very interesting as it allows to point out the most influential factors. The latter should be known accurately before carrying out measurement/prediction comparison. Normally, the total uncertainty in outputs (due to all inputs uncertainties) and uncertainty in measurement (due to the sensors accuracy) are compared. Then, if the intervals overlap during all the period of measurement, the model is deemed to be *valid*. As far as we're concerned, we only take into account the uncertainty on the measured data (indoor dry-bulb temperature). So, if the predicted output doesn't fall in the uncertainty bands, the model is rejected. Then, the source of discrepancy is investigated by analysing the residual (difference between measurement and prediction).

The classical approach consists in searching for inputs/residual correlation. For this purpose, recent works showed that, concerning empirical validation of building thermal simulation programs, spectral analysis tools are useful. In this article, we demonstrate that those tools are sometimes inefficient and can be advantageously completed with time-frequency analysis. Anyway, expertise in experimental design and modelling are necessary to determine whether the cause of disagreement comes from point (1) or (3).

## 2. Application to a real cell test

### *2.1. The real building description*

The survey concerns a real cell test that was erected at University of Reunion Island for experimental validation of building thermal airflow simulation software [3]. After describing the *building* and recalling some model assumptions, we'll apply the methodology previously introduced to *validate* (actually calibrate) our model.

The studied cell test is a cubic-shaped building with a single window on the south wall and a door on the north one. All vertical walls are identical and are composed of cement fibre and polyurethane, the roof is constituted of steel, polyurethane and cement fibre and the floor of concrete slabs, polystyrene and concrete (see table 1 for more details concerning the walls constitution). The building under consideration is highly insulated. Picture 1 shows a picture of the cell test and on the left, the weather station that provides solicitations (inputs) to our model.

*Picture 1 : Picture of the test-cell view from North-West.*

### *2.2. The model description*

A lumped approach [1] is used to represent the building. It is based on the analogy between the Fourier's equation of conduction and Ohm's law. Such a model leads to a system of equations, called state equations, which in the matrix formalism has the following form :

$$C.\dot{T} = A.T + B \qquad (1)$$

where :

$A$ is the state matrix;

$B$ is the solicitations vector;

$C$ is the capacitances matrix;

$T$ is the state vector including the temperature of the lumped elements;

$\dot{T}$ is the derivative of $T$.

In this survey, we consider the electrical/thermal analogy representation of heat transfer by conduction through walls (cf. scheme 1) which consists in discretizing a wall with three nodes per layer.

*Scheme 1 : Wall spatial discretization and representation.*

The thermal capacitances ($C_1,C_2,C_3$) and the thermal conductances ($K_1,K_2,K_3,K_4,K_5,K_6$) of a layer are respectively function of its thickness, specific heat, density ($e,Cp,\rho$) and thickness, conductivity ($e,\lambda$).

### *2.2.1. Assumptions*

Nodal analysis assumes that heat transfer by conduction through the walls are mono-dimensional. Indoor radiant heat transfer is linearized and the radiative exchange coefficients are taken identical for each wall. For outdoor long-wave radiative heat exchange, we use the following model :

$$\varphi_{lwo} = h_{rc}F_{pc}(T_{sky} - T_{so}) + h_{re}F_{pe}(T_{env} - T_{so})$$

$$\text{with } T_{sky} = T_{ao} - 6$$

$$T_{env} = T_{ao}$$

It is not an easy task to measure sky temperature ($T_{sky}$) and models are usually used. Models of this temperature generally take the form of correlations with other parameters. Previous studies (ref. [3]) lead us to the proposed correlation with $T_{ao}$. In the same way, environmental temperature is usually deemed to be equal to outdoor air temperature.

Indoor convective exchange coefficients are assumed constant for each wall whereas outdoor convective coefficients are function of wind speed and direction. For the latter, we choose the correlation of Sturrock [4].

We assume that heat flux under the floor is null. This assumption is reasonable in the studied building since the floor is thermally uncoupled from the ground.

| Type of Walls | Illustration | Materials | Thickness (m) | Conductivity (W/m.K) | Specific heat (J/Kg.K) | Density (kg/m$^3$) |
|---|---|---|---|---|---|---|
| Vertical walls | 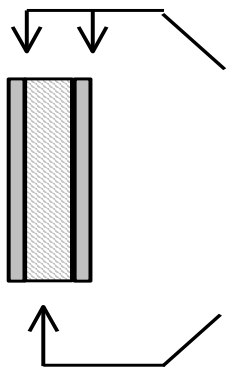 | cement fibre | 0.005 | 0.95 | 1003 | 1600 |
| | | Polyurethane | 0.06 | 0.03 | 1380 | 45 |
| Roof | 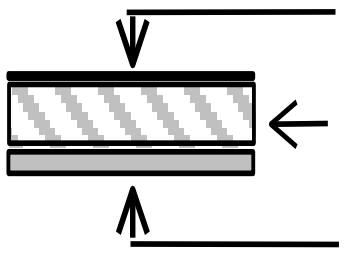 | Sheet steel | 0.003 | 163 | 904 | 2787 |
| | | Polyurethane | 0.05 | see above | " | " |
| | | Cement fibre | see above | " | " | " |
| Floor | 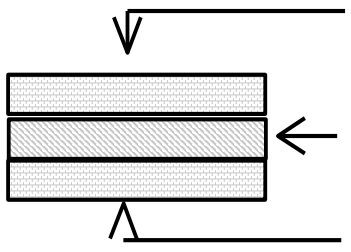 | Concrete slabs | 0.04 | 1.75 | 653 | 2100 |
| | | Polystyrene | 0.04 | 0.04 | 1380 | 25 |
| | | Weight concrete | 0.12 | 1.75 | 653 | 2100 |
| Door | None | Plywood | 0.018 | 0.23 | 1500 | 600 |

*Table 1: Description of the walls of the studied building.*

## 3. Parametric sensitivity analysis

Sensitivity analysis (or SA) of model output is a very important stage in model building and analysis. It is applied in simulation studies in all kinds of disciplines. In building thermal simulation field, SA is more and more applied [5-9]. Indeed, SA can help increase reliability in building thermal simulation software's prediction. We use the method introduced in ref. [9] to identify the most influential factors and evaluate their effect. The proposed method consists in performing several simulation runs by oscillating each parameter sinusoidally over its range of interest. Analysing the spectrum (Fourier transform or power spectral density) of the output, identification of the most influential factors can be easily derived.

We perform 1024 simulations by making each factor vary as a *sinusoid* ranging ±10% with respect to its base value. In the following study, we are looking for the most important parameters upon the predicted indoor air temperature. So, once the simulations are achieved, we calculate the power spectrum density (PSD) of $\Delta T_i = T_{i,base} - T_{i,evol}$

where $T_{i,base}$ is the indoor air temperature obtained with the factors base value at time *i*

and $T_{i,evol}$ the indoor air temperature obtained with the different simulations at time *i*.

As PSD is defined as the spectral decomposition of a signal's variance, we evaluate the influence of each parameter upon the output calculating the amount of the variance of the gaps ($var(\Delta T_i)$) the due to each parameter.

### *3.1. Results*

Though the influence of a parameter depends on the solicitations, we focus only on one day. In effect, the important factors remain the same, whatever the day is, but their contribution changes depending on the inputs. Fig. 1 shows the hourly amount of the variance of $\Delta T$ due to the most influential parameters, more precisely those who explain more than (almost) 1% of $var(\Delta T)$ at a given time. These 11 parameters explain between 80 to 90% of the total variance of the gaps. The remaining amount should be explained by the low effects of the other factors and eventually interactions.

We can note that the influence of the solar window's transmittance is high during the day and progressively decreases during the night. In the contrary, the effects of the variation of the polyurethane's thermal properties (thickness and conductivity) are important during the night and negligible during the day. At last, it is interesting to note that the influence of $h_{rc}$ is constant and is due to the way it's taken into account in the model (see §2.2.1).

The principal influential factors ranked in order of importance are :

- solar transmittance of the south window;
- thickness and thermal conductivity of the polyurethane;
- Outdoor surface radiative heat transfer coefficient with the environment ($h_{rc}$);
- thermal properties of cement fibre;
- thermal properties of concrete slabs on the floor.

*Fig. 1 : Decomposition of the variance of the gaps versus the 11 most important parameters.*

### *3.2. Discussion and physical interpretation*

Physically, ($\rho$ ,$Cp$,$e$) of a material correspond to its thermal capacitance whereas ($e$,$\lambda$) represent its thermal resistance. So, one can notice that it's the thermal inertia of the cement fibre and concrete slabs that have an influence on indoor air temperature. On the other hand, concerning the polyurethane, it's its thermal resistance that has an effect on indoor air temperature. This last results explain why the contribution of the polyurethane's thickness is equal to the one of its conductivity anytime and ditto for the thermal properties of cement fibre and concrete slabs. The latter result corroborates those found by Ezzamari [8] who used an other sensitivity analysis approach to identify the most important factors in a building thermal model.

Window's transmittance is found to be the most important factor. This result is expected as radiation passing through is the first heat source since the cell test's walls are insulated (polyurethane, polystyrene). The preponderance of outdoor radiative heat transfer coefficient with the fictive sky temperature ($h_{rc}$) shows that great care should be taken when outdoor radiative heat transfer is linearized.

## 4. Model's calibration

### *4.1. The experimentation set-up*

An experimentation was conducted during September (fresh season) of 1996. Among others (humidity, surface temperatures,…), two kinds of measurements were made within the cell test : dry-bulb temperature and black globe temperature. Sensors were set at three different heights to take into account air stratification.

*Fig. 2 : Instrumentation of the cell test.*

The aim of this study is to check the ability of our dynamic building simulation program to model the thermal behaviour of the cell test in the simplest configuration. Consequently, the window was set on the south wall so that only diffuse solar radiation got into the room (south hemisphere).

A weather station measured the following climatic solicitations :

- Outdoor dry-bulb temperature;
- Horizontal direct solar radiation;
- Horizontal diffuse solar radiation;
- Relative humidity;
- Wind speed;
- Wind direction.

Data were sampled every minute and averaged every hour. The experimentation lasted fourteen days.

### *4.2. Weather data analysis*

Direct solar radiation is high during the first days whereas the three following days are cloudy as diffuse solar radiation is high (cf. Fig. 3). Wind speed seems to be correlated to direct solar radiation. In effect, when direct solar radiation is high, wind speed increases (cf. 11$^{th}$ et 12$^{th}$ days). Most of the time, the wind comes from east (270°) and corresponds to trade winds.

*Fig. 3 : Weather data.*

### *4.3. Comparison of measured and predicted data*

Black globe temperatures are identical to dry air temperatures (not shown). So, the basis for is taken as the average of the three dry-bulb sensors. The uncertainty on measured temperature is ±0.5°C.

We compare the predicted and measured indoor air temperature after taking off the two first days (because of the warm-up period). Analysis of time variation of the plots shows that there is no delay between measurement and prediction and that the residual is higher during the night (cf. Fig. 4). As the residual doesn't fall in the measurement uncertainty intervals, the model is rejected. The mean and standard deviation of the residual are the following :

| | |
|---|---|
| Mean | 0.267 °C |
| Standard deviation | 0.368 °C |

*Fig. 4 : Comparison of measured / predicted indoor air temperature in the cell test.*

To find the error origins, we first search for correlations between inputs and the residual. Indeed, building thermal simulation programs are generally linear as regard to inputs. As far as our code is concerned, it is the case (see equation (1)). So, when the comparison between measurement and prediction fails, it is expected that the residual (measured - modelled) is correlated to one of the inputs at less. This study is interesting as the information found may help the analyst to search in the good direction to improve his model. Moreover, compared to sensitivity analysis, this study is inexpensive as it requires no more simulation runs.

### *4.4. Inputs / residual correlations survey*

There are different tools and tests to quantify the dependence between two or more signals [10]. For instance, cross-correlation for time dependence or squared coherency in the spectral field. Ramdani & al. [11] recommend spectral analysis to improve building thermal models. In the following, we first attempt to identify correlation between the residuals and the weather data by use of those tools. To exhibit correlation between two signals (for instance inputs / residual), they use a statistical test in the spectral field. Hence, two signals are said significantly correlated if at a given frequency, their squared coherency is greater than an associated threshold called : the non-zero coherency (see ref. [9-10] for more details). Then, a decomposition of prediction error variance in terms of inputs contribution is calculated in order to identify the inputs which are not correctly taken into account. To do so, conditioned signals are estimated. Concerning the latter decomposition, the authors indicate that the less the inputs are correlated the better the variance disaggregation. However, it is important to note that the tools named above require stationary signals.

#### 4.4.1. *Spectral analysis*

The smoothed power spectral density of the residual (cf. Fig. 5) shows that the power of this signal is confined around $0.041h^{-1}$ (the 1 $day^{-1}$ harmonic) and in low frequency. Consequently, we're going to search for the inputs that are correlated to the residual on these frequency ranges. Firstable, let's analyse correlations between inputs. The test of non-zero coherency between outdoor air temperature and direct solar radiation (cf. Fig. 6) shows that these two inputs are significantly correlated around $0.04h^{-1}$ and $0.08h^{-1}$. The other spectrum coherencies give the same information. Hence, it won't be possible to know exactly which one is actually correlated to the residual in these two frequency ranges. The previous result is confirmed while looking at the spectral coherencies between the residual and the inputs (cf. Fig. 7). In fact, according to this figure, only direct solar radiation seems to be correlated to the residual in the low frequency range.

As it was noted previously, the fact that inputs are correlated to one another is a problem. To illustrate so, we perform the calculation of each input contribution to the residual variance. As the authors explain, the decomposition is sensitive to inputs ranking. Fig. 8 et 9 represent the error disaggregation on three different frequency ranges (also on the whole frequency range) for two different inputs ranking. As it is expected, results are different specially on the second frequency range, [$0.02h^{-1}$ $0.06h^{-1}$] which contains the 1 $day^{-1}$ harmonic. In this frequency range, it is impossible to identify the input(s) involved in the model disagreement. Yet, it seems that in low frequency range, [$0h^{-1}$ $0.02h^{-1}$], direct solar radiation is correlated to the residual.

The unsuitability of classical spectral analysis is due to the fact that climatic data are non-stationary signals. In other words, this means even though they have the same spectral components, they may occur at different time [12]. Consequently, time-frequency analysis should bring more information about their dependence (in time and frequency simultaneously).

*Fig. 5 : PSD of the residual.*

*Fig. 6 : Correlation between outdoor air temperature / direct solar radiation.*

*Fig. 7 : Correlations between the residual and the inputs.*

*Fig. 8 : Residual's variance decomposition taking outdoor air temperature as first input.*

*Fig. 9 : Residual's variance decomposition taking direct solar radiation as first input.*

### *4.4.2. Time-frequency analysis : The STFT*

Two-dimensional analysis is a modern technique to extract information from signals. The interest of this kind of tools is that they simultaneously decompose a signal in time and frequency. One can guess that the information obtained are richer than those given in only spectral or time domain analysis. There are numerous techniques which can be regrouped into two families : time-frequency (for instance, the STFT) and time-scale (for instance, the wavelet transform). The aim of this paper is to apply the suitable technique to our data so we won't develop those tools deeply (see ref. [13] for more details).

The first and intuitive time-frequency tool is the short time Fourier transform (STFT). It consists in performing Fourier spectrum through a short sliding window. Time localization is given approximately by the window's position on the signal. The STFT's squared magnitude representation is called a spectrogram. The information obtained with STFT is sufficient here as the spectral content of our data is poor. Indeed, wavelet transform is suitable for multiresolution signal.

To identify correlation around $0.041h^{-1}$ we're going to analyse the STFT of each data. In this study, the main idea consists in checking whether the spectrogram of the residual is similar to the spectrogram of one of the inputs (which means correlation). Actually, to obtain interesting spectrograms, data must be first filtered in low frequency. Indeed, the power of a signal in low frequency may conceal information in higher frequency. Anyway, in our case, filtering is justified as we are specially interested in analysing data around 1 $day^{-1}$ harmonic ($\sim 0.041h^{-1}$). For this purpose, a fourth order Butterworth filter (low-pass) is used to remove (in each signal) frequencies lower than $0.02h^{-1}$. Spectrograms were estimated with the toolbox developed by Auger & al. [14].

As we also mentioned above, there is a trade off between time and frequency resolution depending on the window's width. For the present study, a Hanning's window of 49 *h-width* was chosen for reasonable compromise between temporal and spectral resolution.

Fig. 10 represents :

- the filtered residual in time (plot on the top of the figure)
- the power spectral density of the filtered residual (plot on the left)
- the spectrogram of the filtered residual (in the center). The latter principally exhibits a 1 day$^{-1}$ harmonic temporally localised around [50h, 200h, 250h and 325h] (cf. Fig. 10).

Spectrograms of climatic data show that time-frequency signature of the outdoor air temperature (Fig. 11) is very different of those of the solar radiation. Conversely to classical spectral analysis (see §4.3.1), it is now possible to distinguish outdoor air temperature from the other inputs. Spectrograms of diffuse and direct solar radiation (Fig. 12 & 13) illustrate the complementary of these two inputs. Consequently, it will be difficult to identify which one is really correlated to the residual. Comparison of the spectrograms clearly puts forward a correlation between outdoor air temperature and the residual around the 1 day$^{-1}$ harmonic. So, we can conclude that the error origin comes from the way the model accounts for outdoor air temperature.

*Fig. 10 : Spectrogram of the filtered residual.*

*Fig. 11 : Spectrogram of the filtered outdoor air temperature.*

*Fig. 12 : Spectrogram of the filtered diffuse solar radiation.*

*Fig. 13 : Spectrogram of the filtered direct solar radiation.*

We've just identified the principal input involved in the model disagreement. The following step consists in finding the error origin which is an other problem. One may suppose that the correlation between outdoor air temperature and residual is due to air infiltration in the cell test during experimentation. Yet, our knowledge of the experimental design leads us to reject this hypothesis. In the next section, we verify whether the cause of disagreement between the model and measurement is due to the underestimation of the actual value of a parameter.

### 4.5. *Error origin and model improvement*

#### 4.5.1. *The methodology*

To check whether the error origin comes from the unknown value of a parameter, we proceed as follow :

1) We perform a simulation by changing the value of the tested parameter;
2) We calculate the gaps between the new predicted indoor dry air temperature and the one obtained with the base case model;
3) We search for correlation between the residual (measurement – prediction) and the discrepancy.

Finally, once the parameter identified, we search for the optimal value that gives a residual not correlated to the discrepancy calculated previously. Of course, this approach is correct only if the model is linear as regard to the parameter under consideration.

Sensitivity analysis of the model showed that the most influential factors are principally the window's transmittance and thermal conductance of the polyurethane. As there is obviously no link between outdoor air temperature and transmittance, we start with the polyurethane's thermal properties. *Scatter plots* are used to analyse correlation between residual and the discrepancy due to the parameter variation.

#### 4.5.2. *Application*

Fig. 14 clearly puts forward a linear correlation between the residual and the effect of the polyurethane conductivity (or thickness). This result proves that the error origin in the model comes from an underestimation of this parameter. In fact, as we explained above, the influence of the polyurethane's thickness and conductivity are indistinguishable. Consequently, the problem may also comes from the actual value of the polyurethane's thickness. So, we have measured accurately the polyurethane's thickness (the cell test can be dismantled). Then, we search for the optimal value of the latter using the methodology described above and we found : $\lambda = 0.024$ W.m$^{-1}$.K$^{-1}$.

*Fig. 14 : The effect of the variation of the polyurethane's conductivity is correlated to the residual.*

Comparison to measurements indicates that the new model is better than the previous one (cf. Fig. 15). As the residual falls in the uncertainty intervals of measurements the model was judged accurate enough.

The statistical properties of the new residual are :

| | |
|---|---|
| Mean | 0.028 °C |
| Standard deviation | 0.256 °C |

*Fig. 15 : Comparison between measurement and prediction of the new model.*

Firstable, it must be noticed that the link between the thermal resistance (origin of error) and outdoor air temperature (the input correlated to the residual) is heat transfer by conduction.

The estimated value of the polyurethane's conductivity was first deemed doubtful. Indeed, according to our principal source [15], this material has a conductivity between [0.03 0.045] $W.m^{-1}.K^{-1}$. Fortunately, we found two other sources (cf. [16] and [17]) that confirm our estimation of this conductivity.

As measured data were used to improve the model, the latter can't be considered as validated but as calibrated (see ref. [18]). So, to corroborate our model's adequacy, a new set of measurements is necessary. We have recently carried out an experimentation on the same cell test and we present here the new confrontation.

## 5. Model's corroboration

To validate the previous model, we carried out a new experimentation that lasted twenty two days. Yet, to ensure reproducibility of the results, the experimental set-up were different from the previous one. Firstable, the cell test was moved to a different place (from the North to the South of the island). Moreover, the experimentation was performed during the austral summer (instead of winter) and the data was sampled at 15 min (instead of 1h). At last, the window was turned opaque to keep solar radiation from getting in the room.

### *5.1. The new confrontation*

We use the model identified previously but because of the experimentation we change three things :

- A new component is used to represent the opaque *window;*

- The albedo is changed to 0.3 (instead of 0.2) because of the site configuration;
- Instead of the adiabatic condition under the ground, the temperature was assumed equal to the one measured at 10 cm in the ground (see Fig. 16) under the cell test.

The model is in good agreement with measurement as the residual belongs to ±0.5°C most of the time. Its mean and standard deviation are :

| | |
|---|---|
| Mean | -0.012 °C |
| Standard deviation | 0.340 °C |

*Fig. 16 : Temperature measured in the ground at different depths.*

*Fig. 17 : Comparison of measured and predicted data of the optimal model with a new set of measurements.*

### *5.2. Validation test*

We apply a statistical test to check whether the predicted indoor air temperature and measurement had the same mean and the same variance. We choose the one proposed by Kleijnen & Van Groenendaal [19]. This test consists in regressing the difference between measured and predicted data (residual) on their sum :

$D_i = \gamma_o + \gamma_1 Q_i$

where $D_i$ is the difference between measured and predicted indoor air temperature and $Q_i$ is their sum;

$\gamma_o$ and $\gamma_1$ are the regression coefficients.

The test consists in verifying that $\gamma_o = 0$ and $\gamma_1 = 0$ (null-hypothesis, $H_o$). To test this joint hypothesis $H_o$ simultaneously, according that measured and predicted data are identically and independently normally distributed, the full and reduced sum of square errors must be evaluated :

$$SSE_{full} = \sum_{i=1}^{n} D_i - D_i^*$$

$$SSE_{reduced} = \sum_{i=1}^{n} D_i$$

where $D_i^* = C_o + C_1 Q_i$ and $C_o$, $C_1$ are the ordinary least square estimators of the previous regression.

It can be shown that $F_{2,n-2}$ defined by :

$F_{2,n-2} = [(n-2)/2][SSE_{reduced} - SSE_{full}]/SSE_{full}$ is an $F$-statistic with ($2$, $n$-$2$) degrees of freedom. Application of this test to our data ($n$ = 1920, sampling at ¼ h during twenty days) leads us to determine :

$[(n-2)/2][SSE_{reduced} - SSE_{full}]/SSE_{full} = 3.21$

As $Pr(F_{2,n-2} > 3.21) \approx 0.04$, we accept the null-hypothesis $H_o$ with the probability of 4% to be wrong and we conclude that the model is *valid*.

## 6. Conclusion

We've just performed an empirical validation of a cell test's thermal model. The fact that our first model wasn't accurate enough turned the validation problem into an inverse problem. In validation terminology this is called model calibration. Then, the new model predictions were corroborated with a new set of measurements. Calibration requires sensitivity analysis and research of the inputs which are not correctly taken into account by the model. For the latter, we demonstrated in our study that a time-frequency analysis tool called STFT is suitable to exhibit correlation between residual and inputs. Sensitivity analysis also helped to identify the origin of discrepancy between measurement and the model.

The actual error came from an underestimation of the thermal conductivity of the polyurethane. The latter result also demonstrates how it is important to have reliable documentation on data-sets of thermophysical properties of building materials. To corroborate the adequacy of the model, the latter has been confronted to a second set of measurement. The fact that experimental set-up were different from the previous one, ensured reproducibility of the model in the studied configurations (i.e. when only diffuse solar radiation passes through the window). Next empirical validation works will focus on the ability of the program to take into account global solar radiation in a room and to model the effect of ventilation on indoor air temperature.

## 7. Nomenclature

$\varphi_{lwo}$ Outdoor long-wave heat flux radiation density (W.m$^{-2}$)

| | |
|---|---|
| $T_{sky}$ | Fictive sky temperature (K) |
| $T_{env}$ | Fictive environment temperature (K) |
| $T_{ao}$ | Outdoor wall (or window) surface temperature (K) |
| $T_{ao}$ | Outdoor dry-bulb temperature (K) |
| $\sigma^2_x$ or $var(x)$ | Variance of variable $x$ |
| $C_i$ | Node thermal capacitance ($J.m^{-2}.K^{-1}$) |
| $K_i$ | Node thermal conductance ($W.m^{-2}.K^{-1}$) |
| $h_{rc}$ | Outdoor surface radiative heat transfer coefficient with fictive sky temperature ($W.m^{-2}.K^{-1}$) |
| $h_{re}$ | Outdoor surface radiative heat transfer coefficient with the environment ($W.m^{-2}.K^{-1}$) |
| $F_{pc}$ & $F_{pe}$ | View factor between a wall and respectively sky and environment |
| $e$ | Thickness (m) |
| $\lambda$ | Conductivity ($W.m^{-1}.K^{-1}$) |
| $\rho$ | Density ($Kg.m^{-3}$) |
| $Cp$ | Specific heat ($J.Kg^{-1}.K^{-1}$) |
| $\tau$ | Solar transmittance |
| $\alpha_o$ | Outdoor surface absorptance of short-wave radiation |

**Acknowledgements**

The financial contribution of *Conseil Régional de La Réunion* and of Reunion Island delegation of ADEME (Agence Départementale de l'Environnement et de la Maîtrise de l'Energie) to this study is gratefully acknowledged.

## 8. Références

## List of Figure Captions

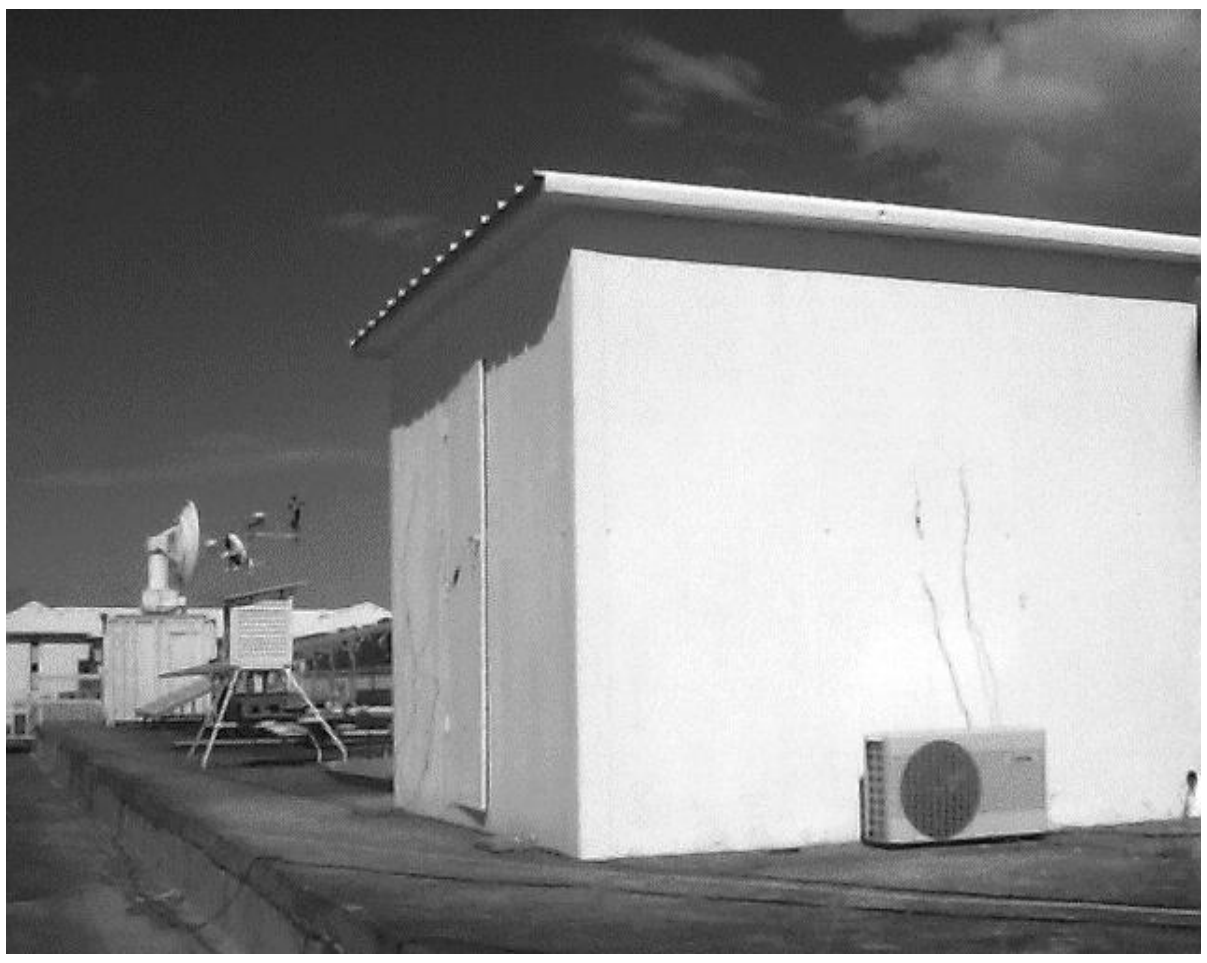

*Picture 1 : Picture of the test-cell view from North-West.*

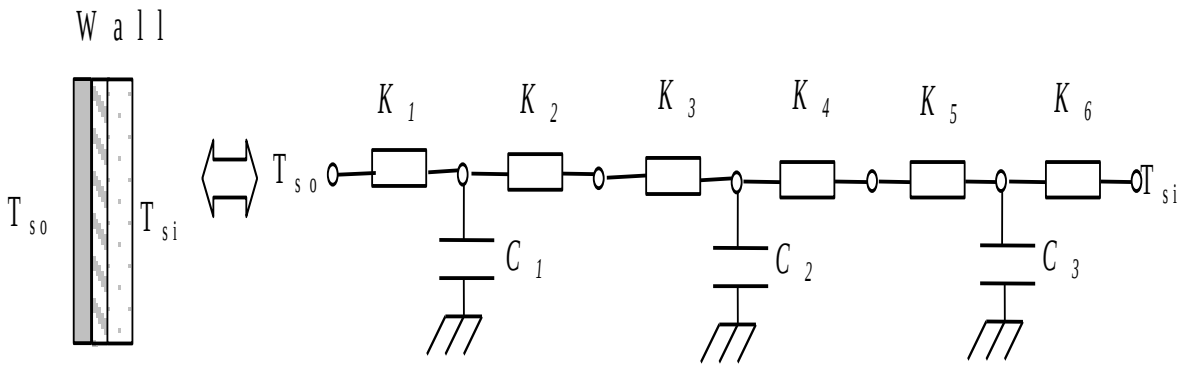

*Scheme 1 : Wall spatial discretization and representation.*

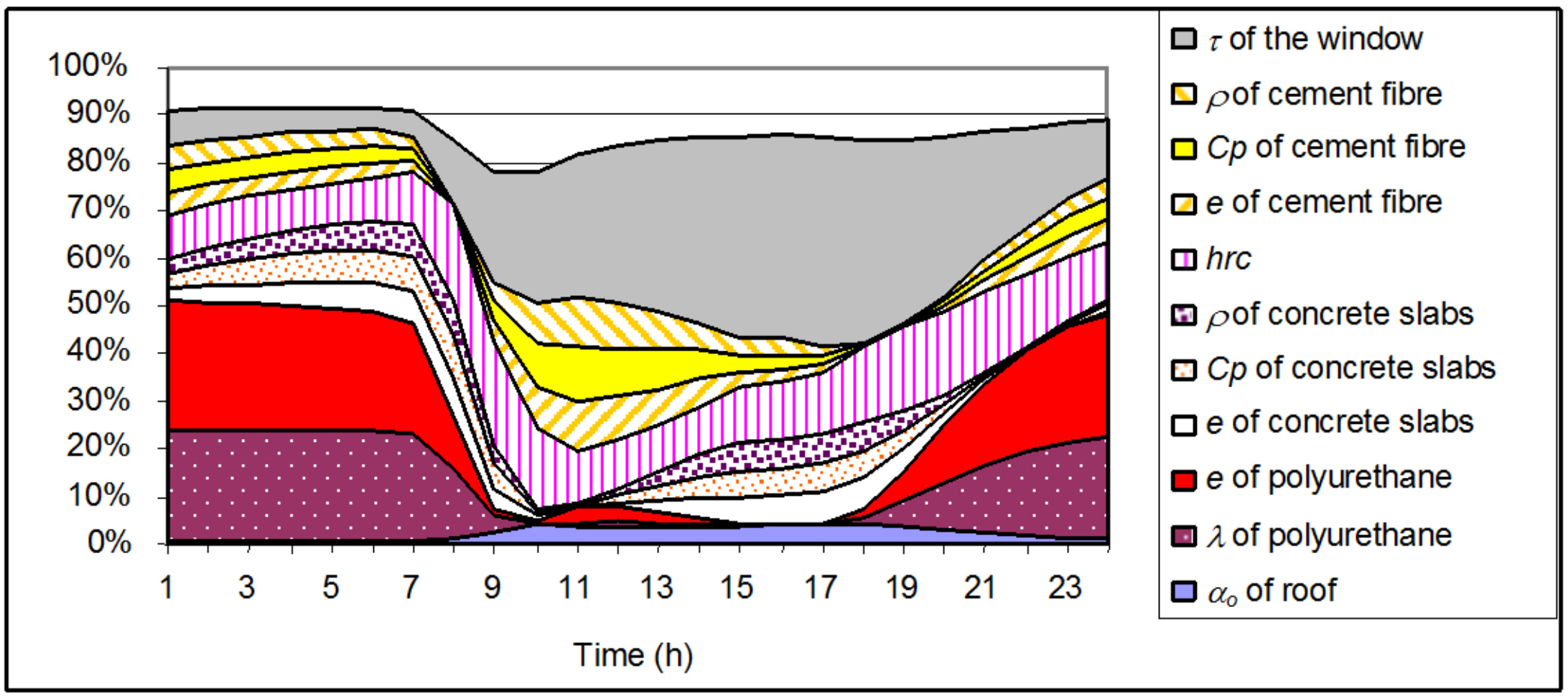

*Fig. 1 : Decomposition of the variance of the gaps versus the 11 most important parameters.*

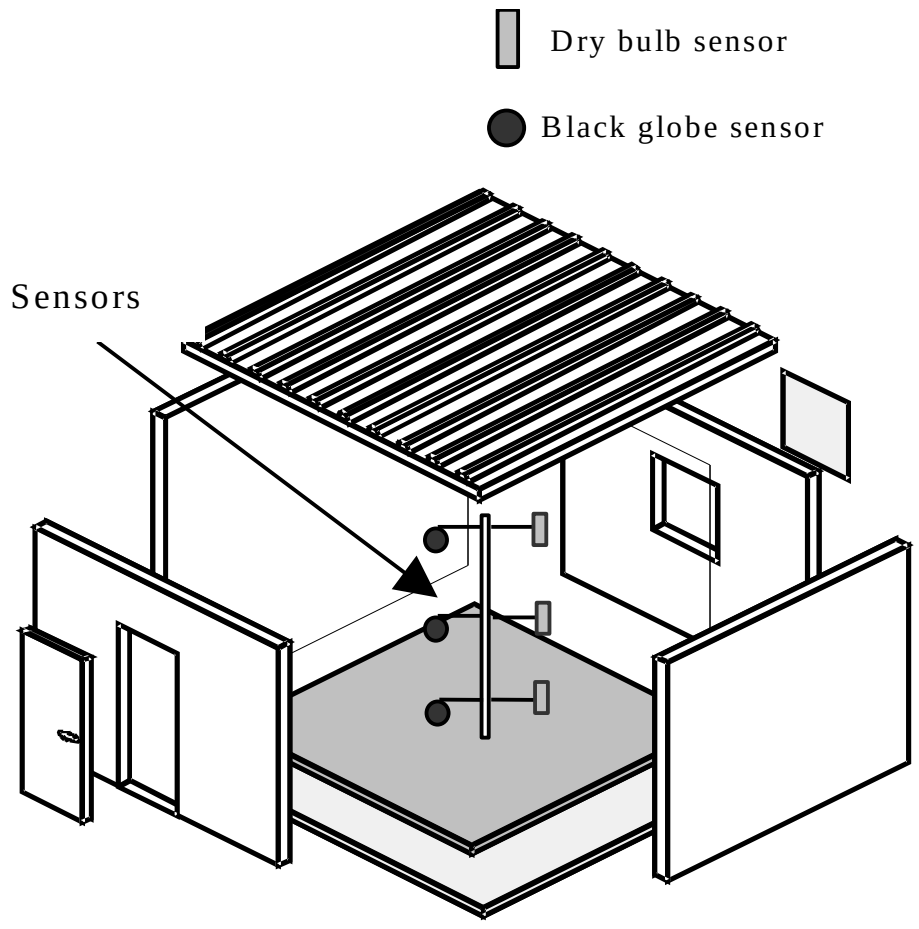

*Fig. 2 : Instrumentation of the cell test.*

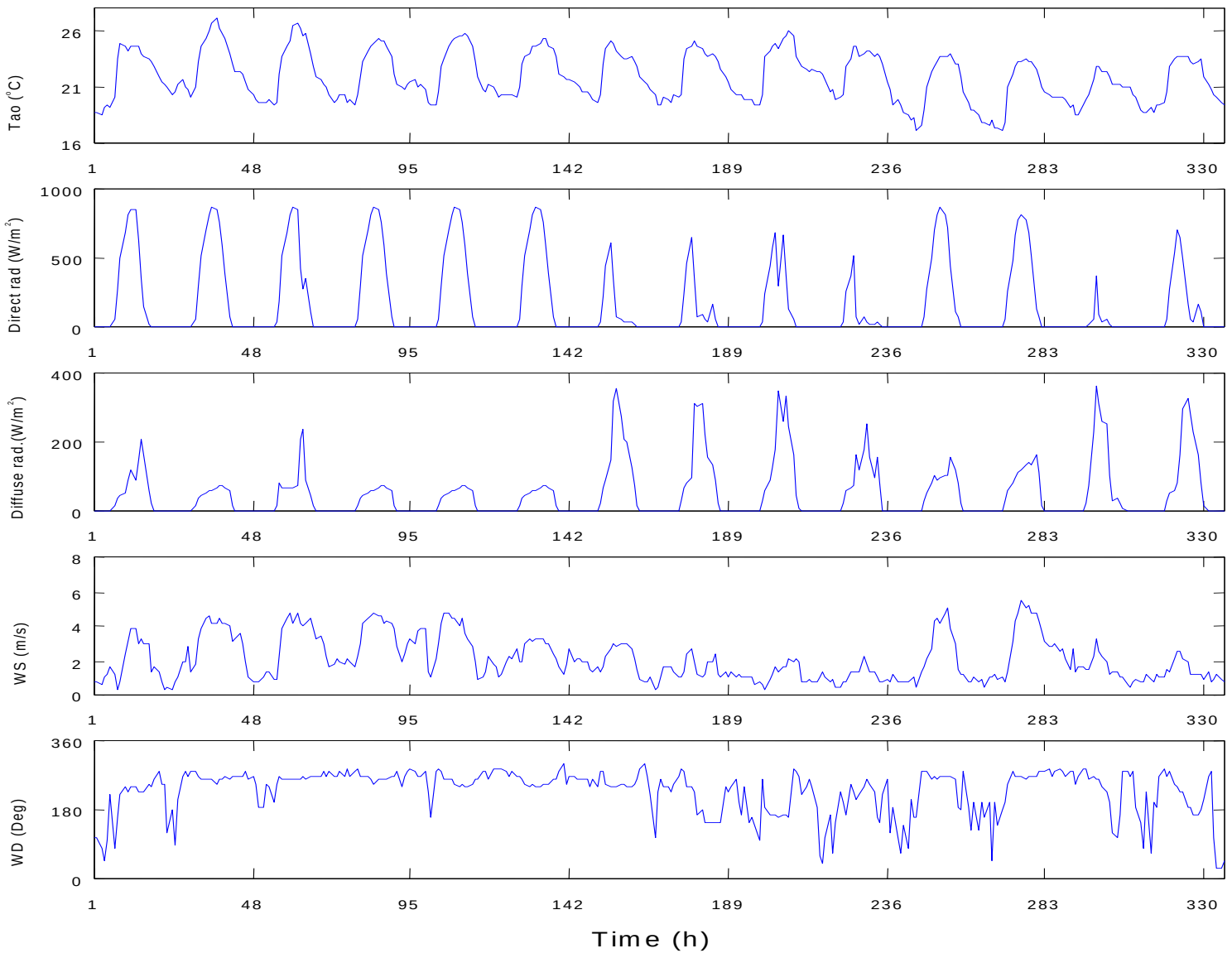

*Fig. 3 : Weather data.*

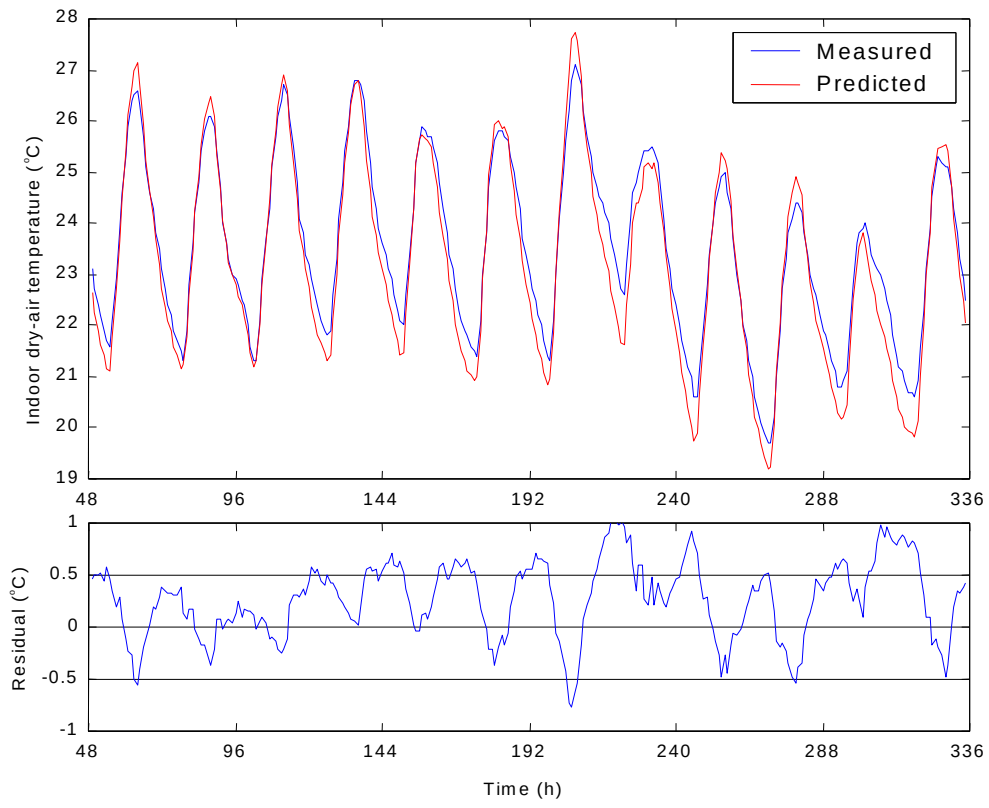

*Fig. 4 : Comparison of measured / predicted indoor air temperature in the cell test.*

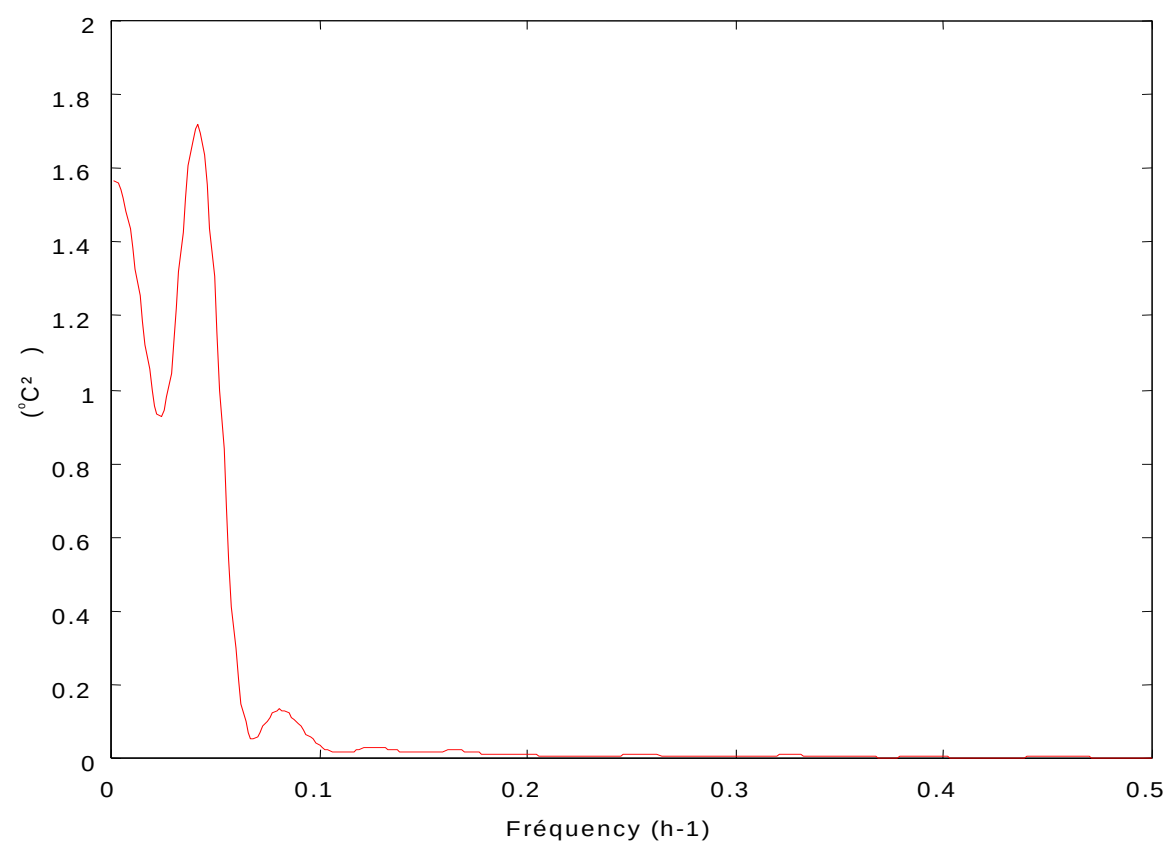

*Fig. 5 : PSD of the residual.*

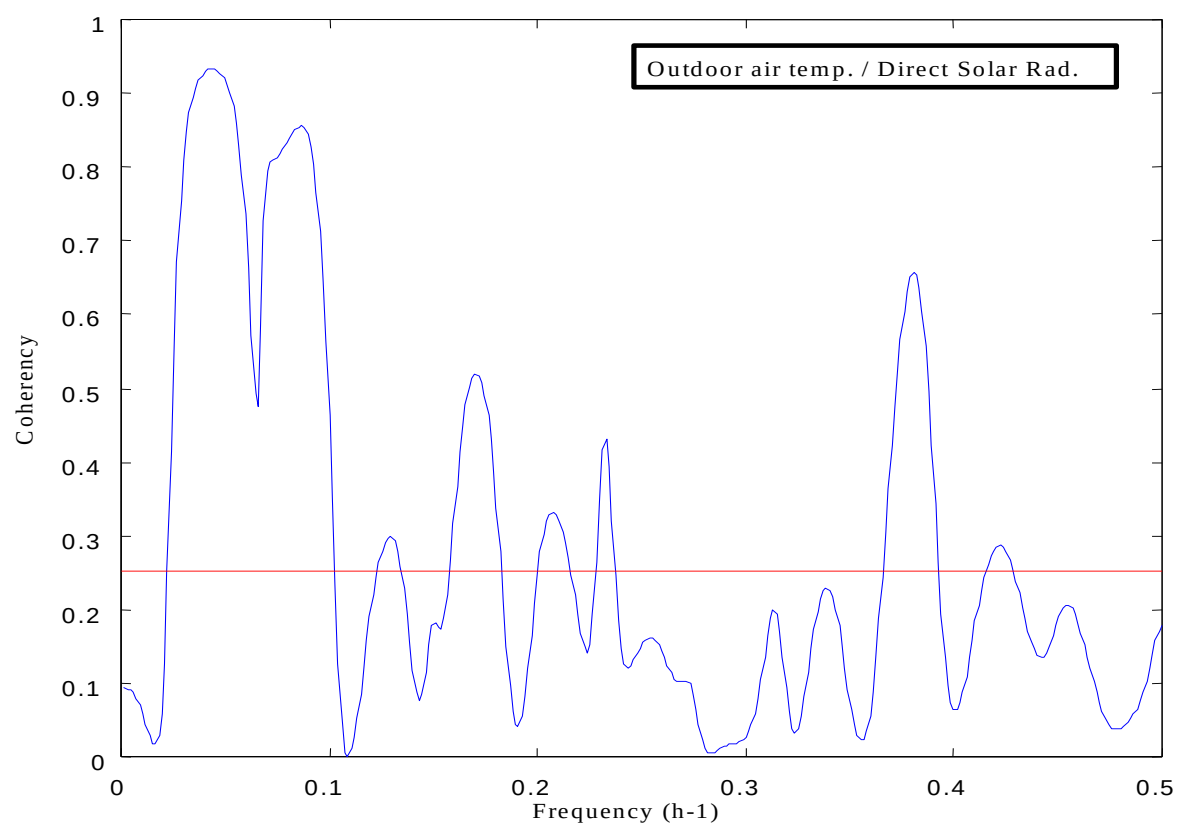

*Fig. 6 : Correlation between outdoor air temperature / direct solar radiation.*

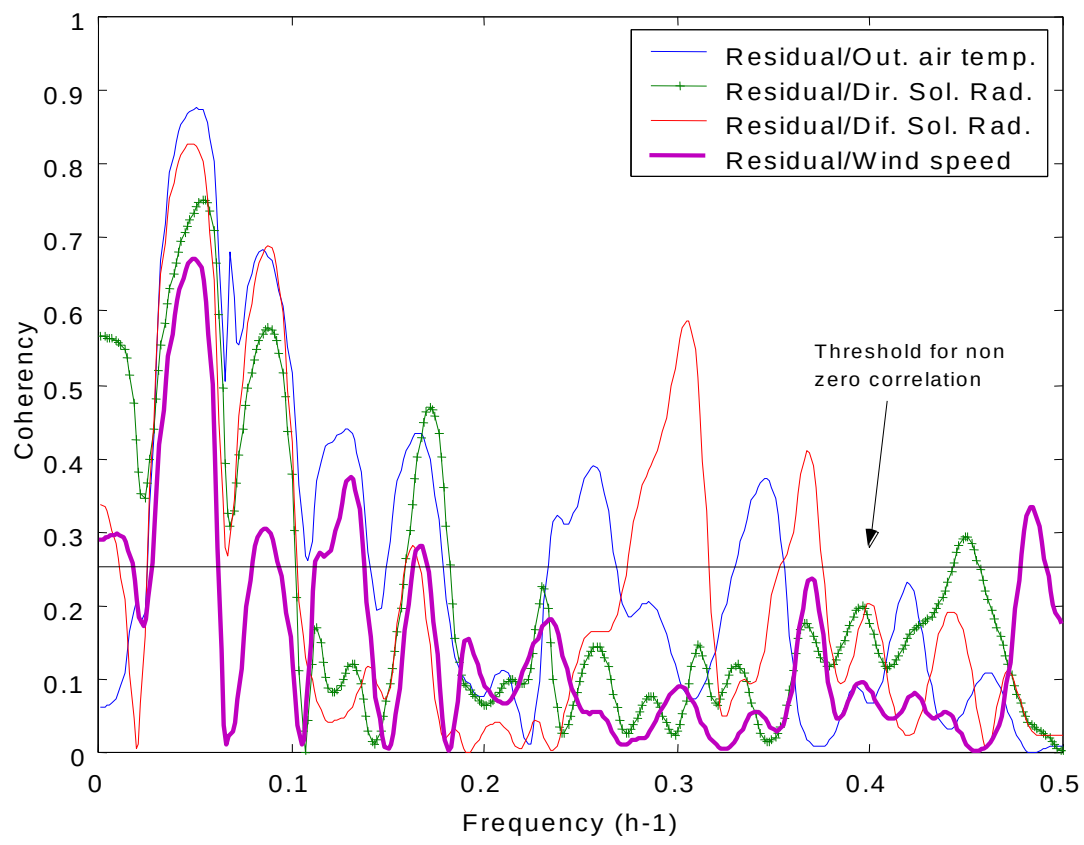

*Fig. 7 : Correlations between the residual and the inputs.*

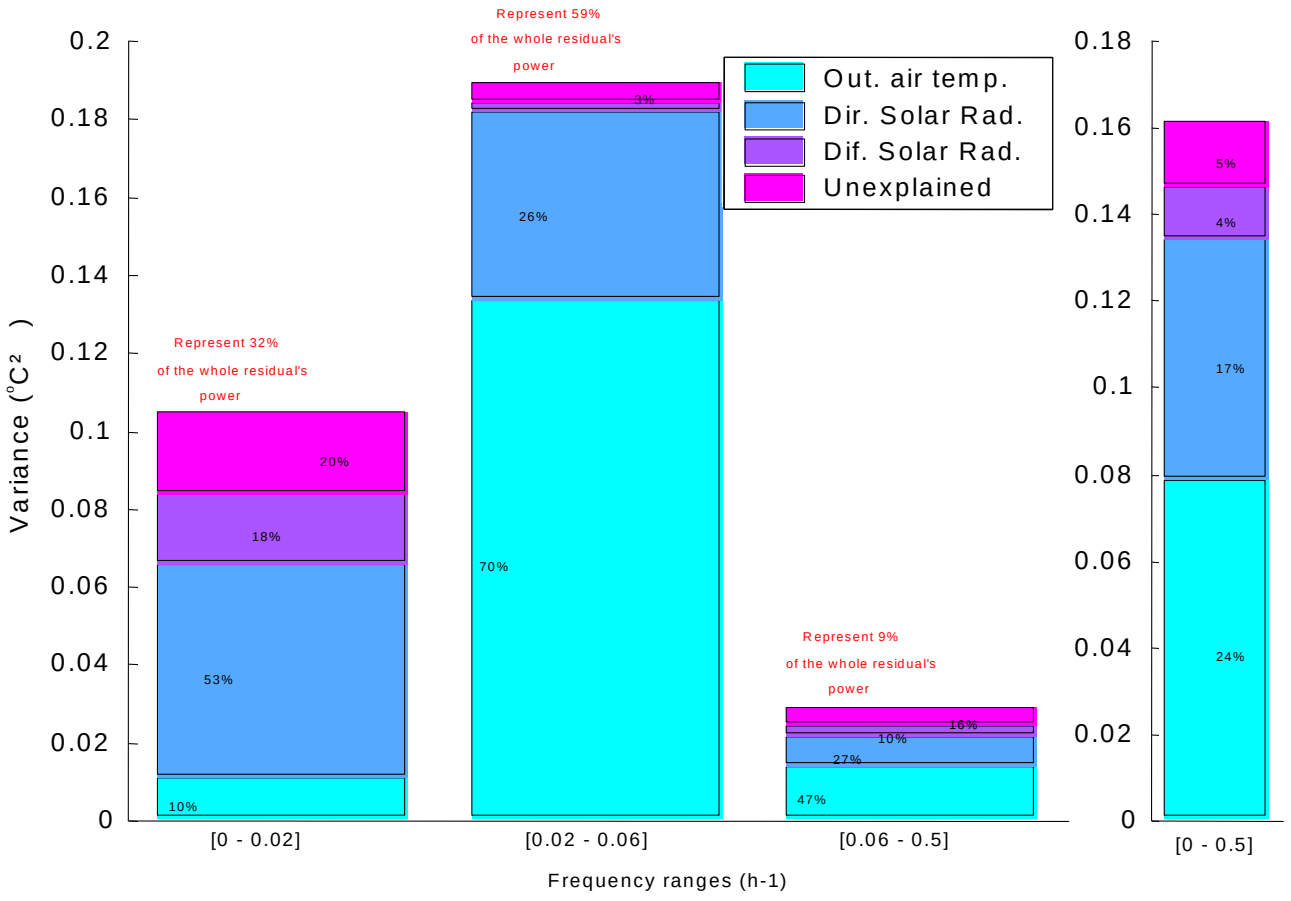

*Fig. 8 : Residual's variance decomposition taking outdoor air temperature as first input.*

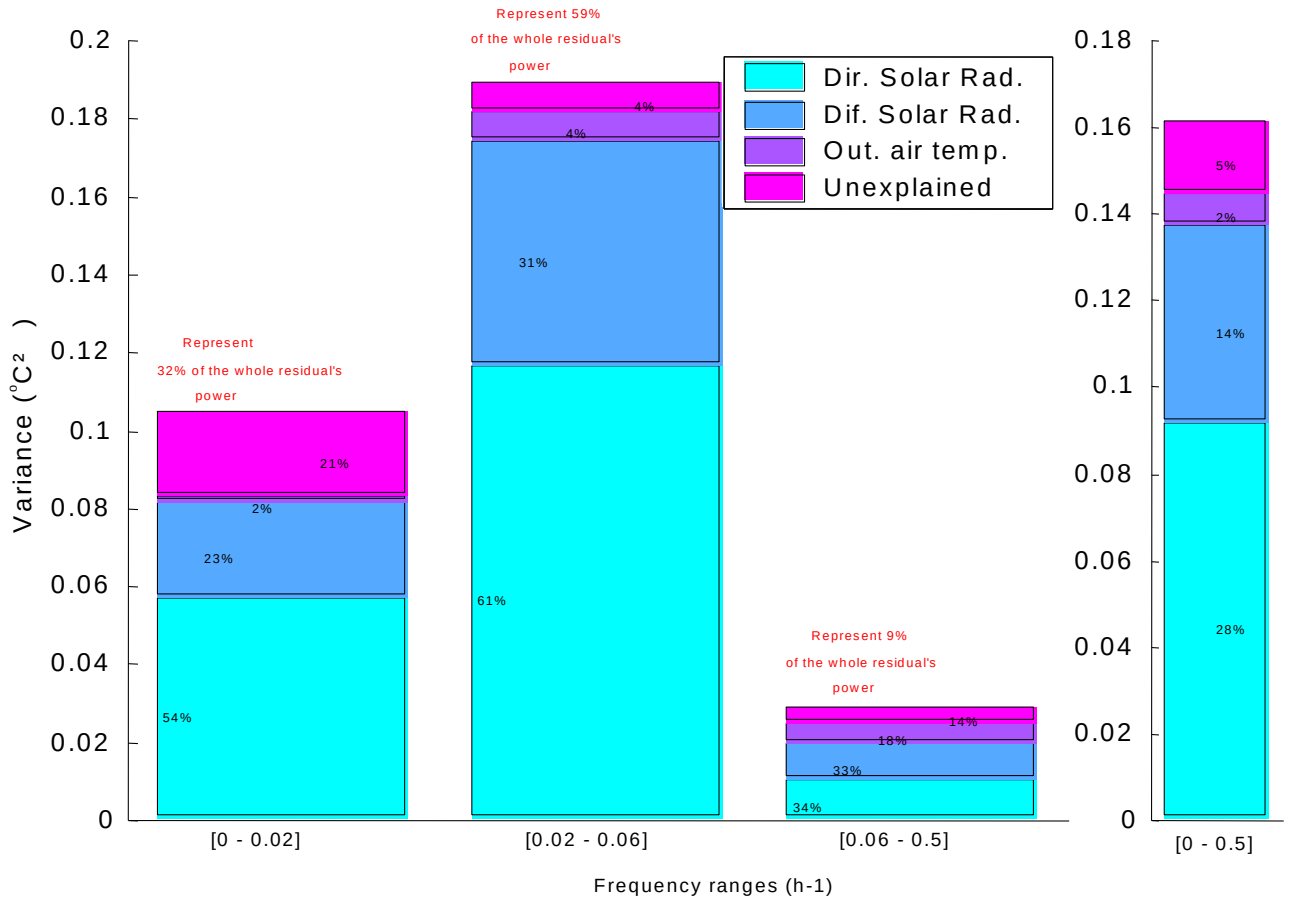

*Fig. 9 : Residual's variance decomposition taking direct solar radiation as first input.*

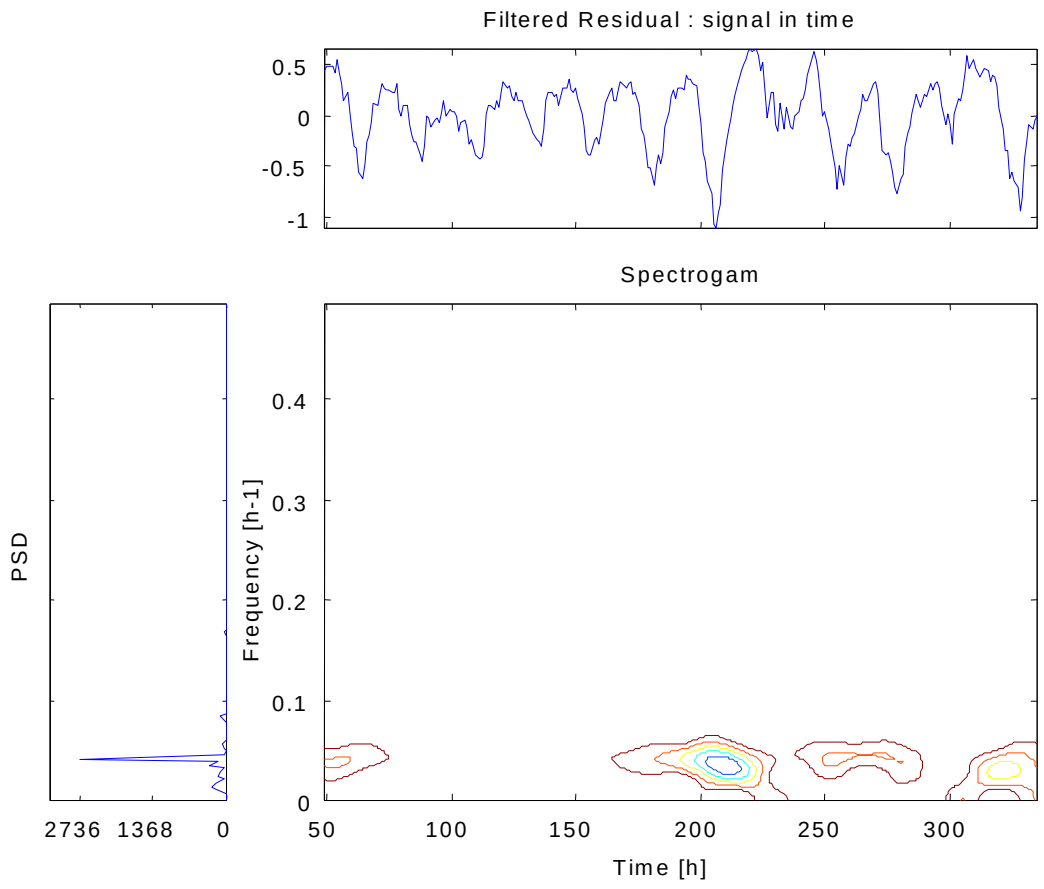

*Fig. 10 : Spectrogram of the filtered residual.*

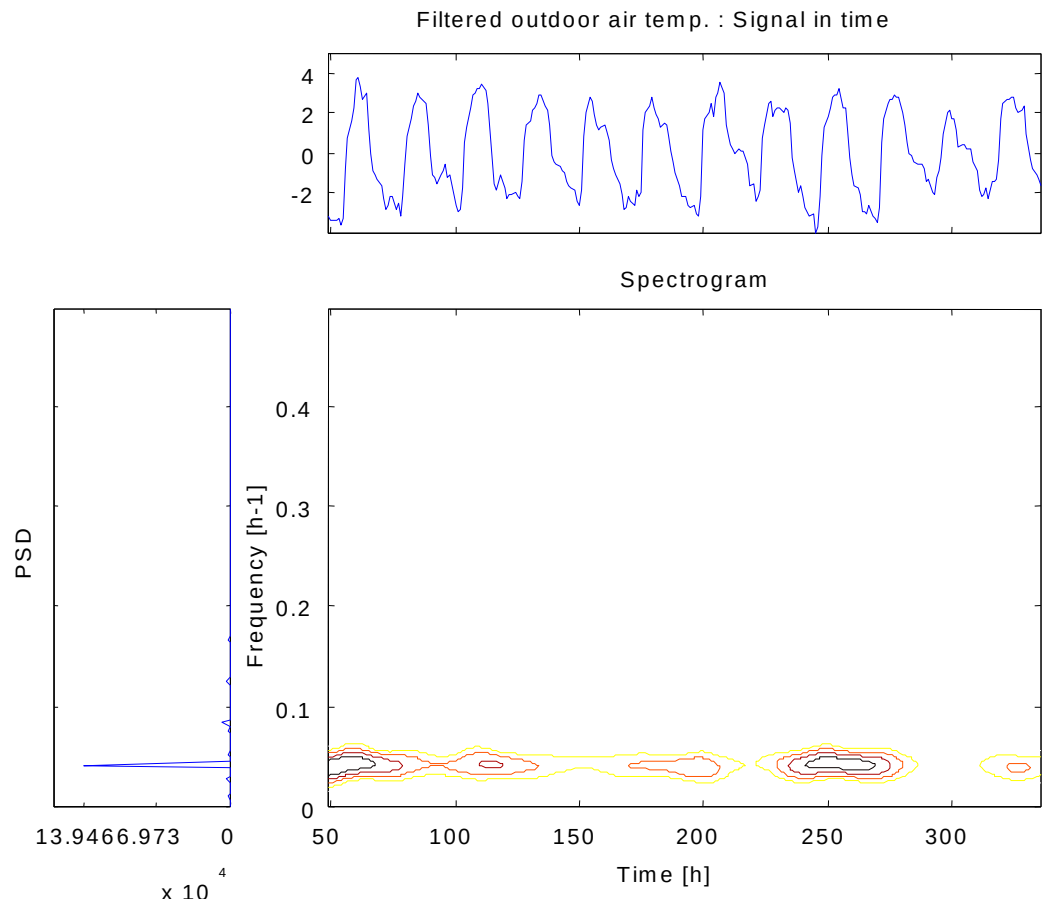

*Fig. 11 : Spectrogram of the filtered outdoor air temperature.*

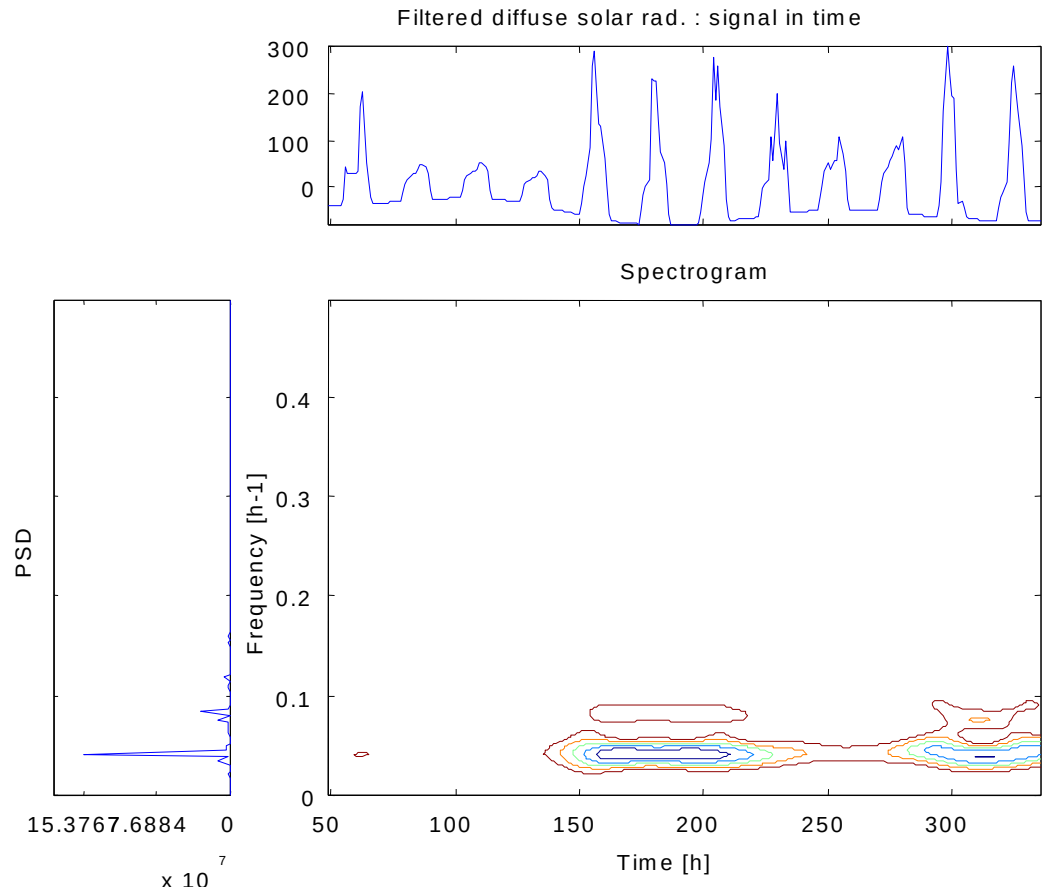

*Fig. 12 : Spectrogram of the filtered diffuse solar radiation.*

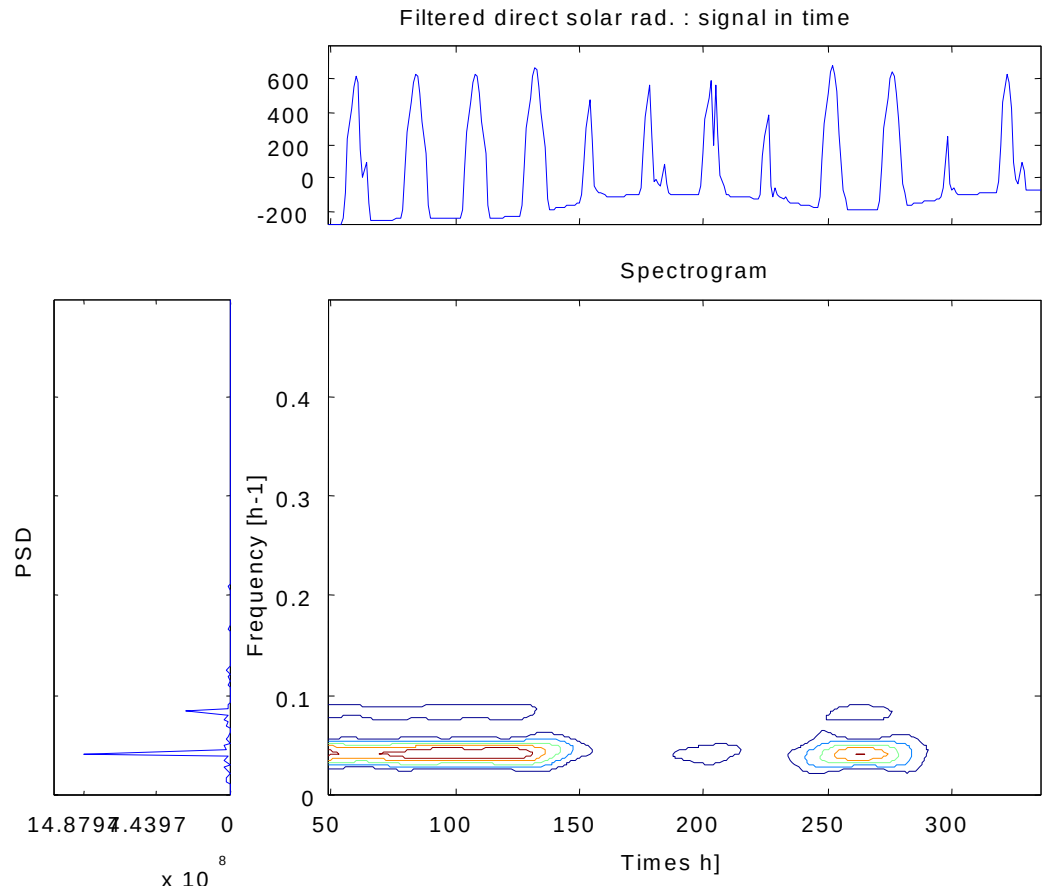

*Fig. 13 : Spectrogram of the filtered direct solar radiation.*

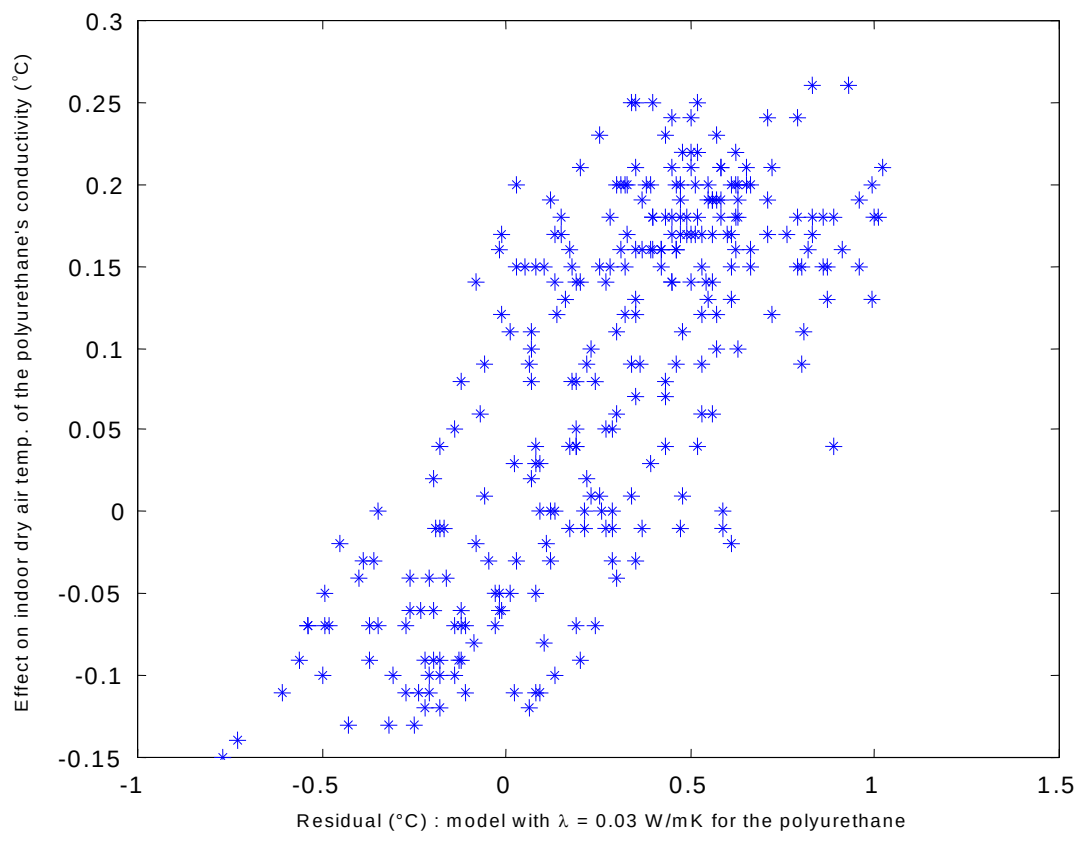

*Fig. 14 : The effect of the variation of the polyurethane's conductivity is correlated to the residual.*

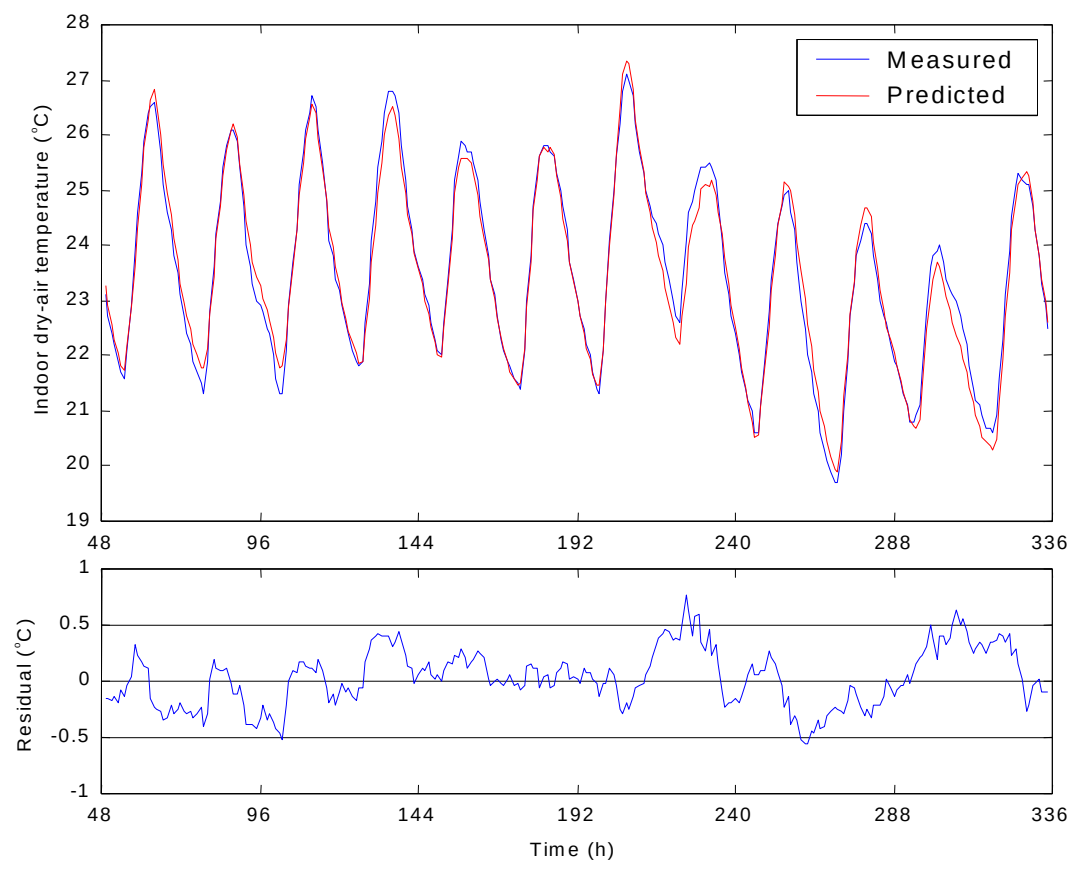

*Fig. 15 : Comparison between measurement and prediction of the new model.*

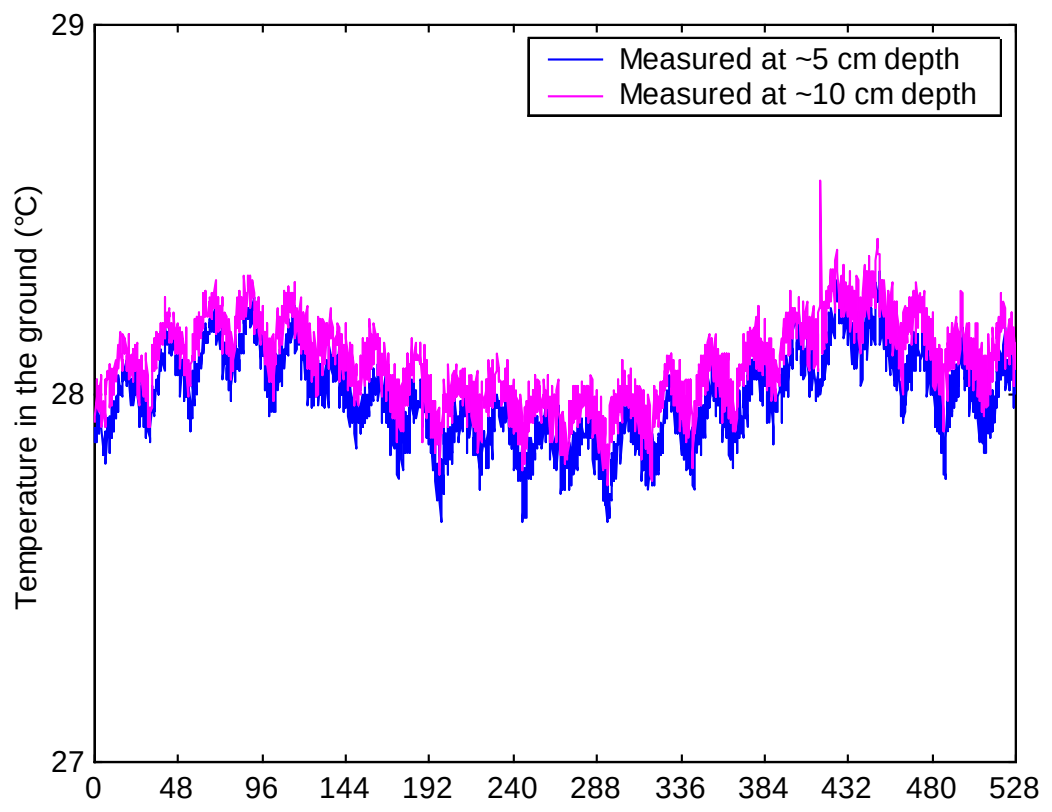

*Fig. 16 : Temperatures measured in the ground at different depths.*

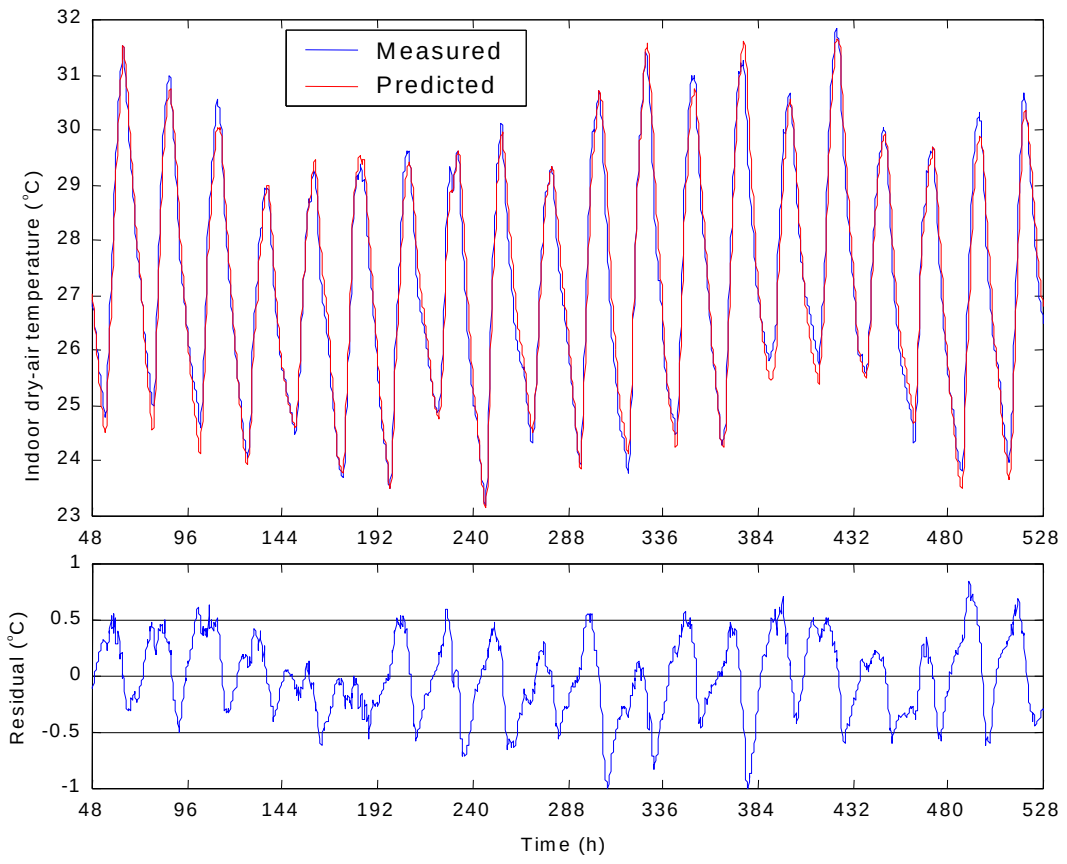

*Fig. 17 : Comparison of measured and predicted data of the optimal model with a new set of measurements.*